# Patient-Specific CT Doses Using DL-based Image Segmentation and GPU-based Monte Carlo Calculations for 10,281 Subjects

Zirui Ye[1], Bei Yao[2], Haoran Zheng[2], Li Tao[1], Ripeng Wang[1], Yankui Chang[1], Zhi Chen[1], Yingming Zhao[2], Wei Wei[2], Xie George Xu[1*]

[1] School of Nuclear Science and Technology, University of Science and Technology of China, Hefei 230026, China
[2] The First Affiliated Hospital, University of Science and Technology of China, Hefei 230001, China

Corresponding author:
Professor Xie George Xu, Ph.D. FAIMBE, FAAPM, FHPS, FANS
University of Science and Technology of China
Huangshan Road 443, Hefei 230026, Anhui, China
Email: xgxu@ustc.edu.cn

# Abstract

**Background:** Computed tomography (CT) scans are a major source of medical radiation exposure worldwide. In countries like China, the frequency of CT scans has grown rapidly, particularly in routine physical examinations where chest CT scans are increasingly common. Accurate estimation of organ doses is crucial for assessing radiation risk and optimizing imaging protocols. However, traditional methods face challenges due to the labor-intensive process of manual organ segmentation and the computational demands of Monte Carlo (MC) dose calculations.

**Purpose:** In this study, we present a novel method that combines automatic image segmentation with GPU-accelerated MC simulations to compute patient-specific organ doses for a large cohort of 10,281 individuals (6419 males and 3862 females) undergoing CT examinations for physical examinations at a Chinese hospital. This is the first big-data study of its kind involving such a large population for CT dosimetry.

**Methods:** Our workflow involved three key steps. First, we collected and anonymized CT images and health metrics (age, gender, height, weight) of the subjects from the hospital's database. Second, we utilized DeepContour, a segmentation tool based on deep learning (DL), to automatically segment organs from the CT images. Additionally, we performed GPU-accelerated rapid MC organ dose calculations using a validated scanner model and ARCEHR-CT. Third, we conducted comprehensive statistical analysis of doses for nine organs: esophagus, heart, lungs, liver, pancreas, spleen, stomach, breasts, and spinal cord.

**Results:** The results show considerable inter-individual variability in $CTDI_{vol}$-normalized organ doses, even among subjects with similar body mass index (BMI) or water equivalent diameter (WED). Patient-specific organ doses vary widely, ranging from 33% to 164% normalized by the doses from International Commission on Radiological Protection (ICRP) Adult Reference Phantoms. Statistical analyses indicate that the "Reference Man" based average phantoms can lead to significant dosimetric uncertainties, with relative errors exceeding 50% in some cases. These findings underscore the fact that previous assessments of radiation risk may be inaccurate. Statistical analysis reveals strong correlations between organ doses and health metrics, including weight, BMI, WED, and Size-Specific Dose Estimate (SSDE), suggesting that these factors can serve as surrogate for dose estimation. It took our computational tool, on average, 135 seconds per subject, using a single NVIDIA RTX 3080 GPU card.

**Conclusions:** By processing a total of 10,281 subjects undergoing CT examination in a busy Chinese hospital, this study has demonstrated the feasibility and cost-effectiveness of combining DL-based multi-organ segmentation with GPU-accelerated MC simulations for large-scale, patient-specific CT dosimetry. The big-data analysis provides interesting data for improving CT dosimetry and risk assessment by avoiding uncertainties that were neglected in the past.

# 1. Introduction

Among nontherapeutic medical radiation sources, computed tomography (CT) is the main contributor to the worldwide collective radiation dose[1,2]. In China, where air quality has been a health concern, the checklist for routine physical examinations has increasingly included chest X-ray CT scans. Although the number of CT scans per capita in China is still relatively low today in comparison with that in many developed countries, the total number of medically exposed Chinese individuals (to CT and other X-ray related procedures) is increasing rapidly, raising questions about the necessity, potential radiation risk and cost. In Shanghai — the most populated Chinese city, the annual frequency of CT examinations in 2016 reached 304 per thousand people which is 2.74 times the annual frequency of 2007[3]. The most recent epidemiological analysis indicated that, in a cohort of 10,000 children who underwent CT scans with an average exposure of 8 mGy, approximately 1 to 2 individuals may develop hematological malignancies attributable to this radiation within the following 12 years[4]. Worldwide, there has been a strong desire to better estimate and manage radiation risk from CT examinations. Although it has been suggested that radiation risk assessment should be based on radiation dose delivered to the exposed organs or tissues[5], there is little data on patient-specific CT organ doses for large population sizes, owing partially to the lack of computational tools.

Monte Carlo (MC) calculations, combined with CT scanner models and patient-specific phantoms have been used for CT dosimetry research for some time. However, two challenges limit the method's routine and wider clinical application: On the one hand, manual delineation of multiple radiosensitive organs or tissues from CT images of a large number of individual subjects is prohibitively laborious and time-consuming. On the other hand, MC simulations, known as the "gold standard" of dose calculation method, are computationally expensive and slow due to the statistical nature[6].

In recent years, automatic multi-organ segmentation methods based on deep learning (DL)[7–9] and fast MC dose computing methods based on graphics processing unit (GPU) devices have been reported for radiotherapy dosimetry applications[10–12]. In the field of CT dosimetry, two groups have explored the feasibility of patient-specific organ-level dose estimation for CT by combining automatic segmentation and GPU-based MC tools[13,14]. To date, however, there has been no demonstration of such modern computational tools for a cohort consisting of more than 100 subjects and, as a result, the full potential of big-data analysis remains to be seen.

This study proposes a novel method that computes and analyzes patient-specific organ doses for a cohort of 10,281 heathy individuals undergoing CT examinations in a Chinese hospital. To our knowledge, this is the first of such big-data study. This study is carried out using a unique computational platform consisting of an automatic segmentation tool for constructing patient-specific phantoms, a GPU-based rapid MC dose computing tool, and a comprehensive data analytical tool. Finally, in light of rapidly increasing trend in CT utility in China, we are interested in showing the feasibility of an accurate and fast method deployable in routine hospital workflows.

## 2. Materials and Methods

The workflow is illustrated in Figure 1. This study includes three essential steps:

(1) We collected a large dataset comprising 10,281 subjects from the First Affiliated Hospital of the University of Science and Technology of China.

(2) We calculated the organ doses of our proposed method, and compared them with the results of the ICRP Adult Reference Phantoms. In this step, we integrated the commercial software, DeepContour[14–16], consisting of a convolutional neural network (CNN) designed for automatic organ segmentation from CT image. In addition, we employed the GPU-accelerated MC dose calculation software ARCHER-CT[11] for simulating X-ray transport and dosimetry, in combination with a experimentally validated CT scanner model[17–19].

(3) We performed a detailed statistical analysis of the subjects' organ doses along with other health metrics.

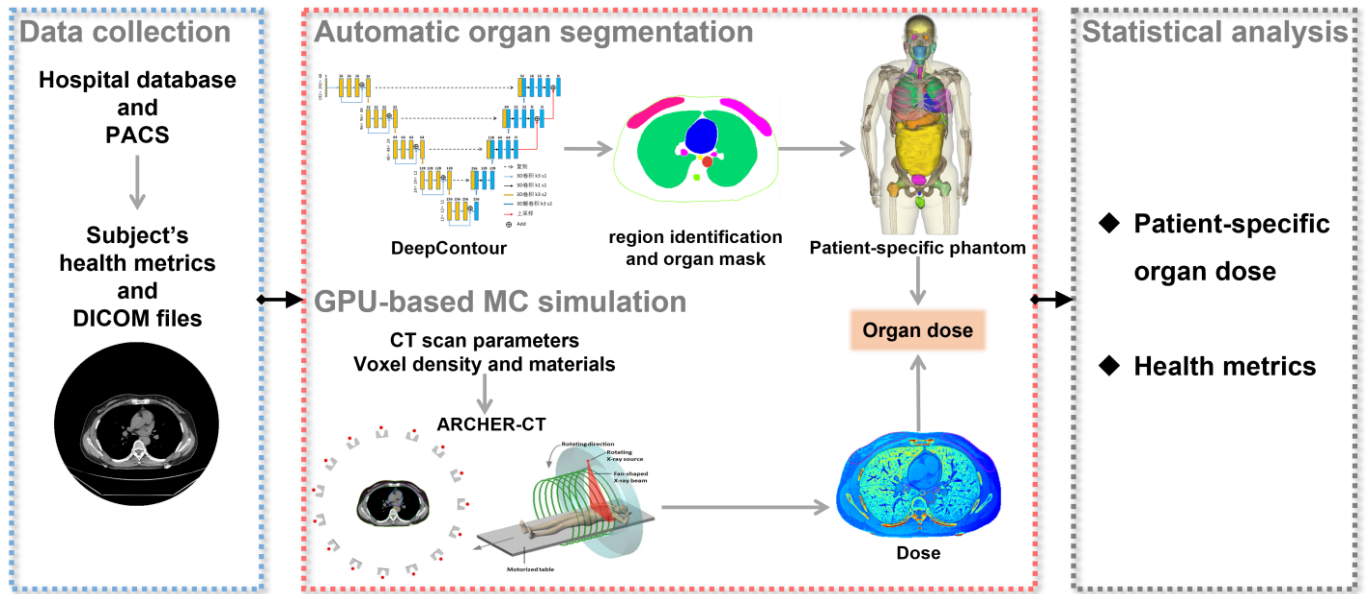

**Figure 1.** Flow chart of the CT dose computation and analysis software platform. First, the subject's data is collected. Second, CT images are automatically segmented, before the corresponding patient-specific phantom is constructed. Then, dose to the phantom is simulated rapidly. Finally, statistical analysis of organ doses and other metrics is conducted.

## 2.1 Data collection

With the approval of the Ethics Committee of the First Affiliated Hospital of the University of Science and Technology of China, this study conducted a retrospective analysis of randomly collected physical examination data from 10,281 CT scan subjects at the Health Management Center of the hospital. Data for all subjects was anonymized before information including age at exposure, gender, height, weight are extracted from the hospital information system (HIS) database using Structured Query Language (SQL). The CT image data taken during the subjects' physical examinations were acquired in bulk, through the Digital Imaging and Communications in Medicine (DICOM) protocol from the hospital's Picture Archiving and Communication System (PACS) network interface. Subsequently, only images and the parameters of the CT scanner were extracted from DICOM files.

Initially, data from 10,500 subjects over the past five years was collected. Out of these, 81 cases were excluded due to duplicated CT examinations and 86 cases were excluded due to at least one missing contours of the nine organs of interest in this study. Moreover, only 52 cases included Automatic Tube Current Modulation (ATCM); these cases were excluded due to insufficient sample size for data analysis. After applying these exclusion criteria, a total of 10,281 subjects were retained for the final analysis.

The range of these CT scans extends from the seventh cervical vertebra (C7) to the second lumbar vertebra (L3), covering the lower neck, thorax, and upper abdomen. As this is a routine physical examination, the scan ranges were consistent for all subjects, most of whom were healthy without significant anatomical alterations from malignant diseases. None of the subjects received any contrast agents. For all DICOM CT image slices, a pixel resolution of 512×512 was maintained. The mean (± standard deviation) pixel size was 0.81 (± 0.04) mm. For each subject, the CT scan from a single examination was reconstructed into two series of images with the slice thicknesses of 1.25 mm and 6 mm, respectively. To ensure accuracy, this study focuses exclusively on the results from images of the 1.25-mm slice thickness. No resampling was performed during the data processing. In other words, the original DICOM files were used directly as inputs to the codes. The bed was removed in each slice of the CT images for all subjects.

## 2.2 Development of a CT organ dose calculation software platform

To enable large-scale computations, we developed a preliminary CT dose computation and analysis software platform based on the dosimetry engine and implemented it in Python Version: 3.11. As illustrated in Figure 1, after the subject's CT images were automatically segmented and the corresponding patient-specific phantom was constructed based on the physical properties (i.e., attenuation, composition and shape), we performed a GPU-based rapid dose computation on the phantom. This approach allowed us to obtain patient-specific 3D dose distributions and organ doses normalized by volume CT Dose Index ($CTDI_{vol}$).

## 2.2.1 Automatic multi-organ segmentation

To construct patient-specific phantoms for individual subjects, it is necessary to extract 3D matrices from the CT images that capture the physical properties, including attenuation, composition, and shape. Then organ contours must be delineated as masks within voxel-to-voxel matrices of the same dimensions. This allows for the calculation of organ mass from the density matrix and organ dose from the dose matrix. Thus, we employed the automatic organ segmentation software, DeepContour (V1.0), based on the 3D ResU-Net architecture. The accuracy of the segmentations produced by an early version of this software (formerly known as DeepViewer) has been validated in our previous studies[14–16]. Specifically, DeepContour supports the automatic identification of scanned regions and the automatic segmentation of more than 80 organs across various regions, including the head and neck, thorax, abdomen, and pelvis. Compared to segmentation results obtained by experienced clinicians, for most organs throughout the body, the Dice similarity coefficient (DSC) reached a value of 0.9. For the current version, the DSC of major organs, such as the brain, lungs, liver, and bladder, even surpassed 0.95[20].

In this study, the automatic segmentation results for the CT images of the 10,281 subjects included contours for 17 organs: body, spinal cord, spinal cavity, trachea, lungs, esophagus, heart, aorta, breasts, liver, spleen, stomach, duodenum, pancreas, kidneys, bowel, and small intestine. For brevity, our analysis considered only the doses for the nine organs that are typically the organs of focus when considering radiation protection: esophagus, heart, lungs, liver, pancreas, spleen, stomach, breasts, and spinal cord.

Based on the organ segmentation masks, additional information was derived. First, by analyzing the range of slices in which an organ's mask appears, we can identify the slices corresponding to that organ, such as the lung region, liver region, or heart region. Second, using the body mask, we can calculate the WED of each CT slice. The WED, as suggested by the American Association of Physicists in Medicine (AAPM), can describe the body size of a subject more accurately than can other indices and can be utilized for calculating the Size-Specific Dose Estimates (SSDEs)[21]. The SSDE is calculated by converting the $CTDI_{vol}$ measured using 32 cm diameter polymethyl methacrylate (PMMA) $CTDI_{vol}$ phantoms to a specific adult subject's body size. The WED of each CT slice is derived from the CT values within the body region of the CT slice image, with the specific formula as follows:

$$WED = 2\sqrt{\left(\frac{\overline{CT_{body}}}{1000}+1\right)\frac{A_{body}}{\pi}} \quad (1)$$
$$f_{size}(WED) = 3.704 \times e^{-0.0367 \times WED} \quad (2)$$
$$SSDE = f_{size}(WED) \times CTDI_{vol} \quad (3)$$

where $\overline{CT_{body}}$ represents the average CT value of all pixels within the body region of the CT slice image, and $A_{body}$ denotes the total area of the body region of the CT slice image. $f_{size}(WED)$ is the SSDE conversion factor.

For a CT scan with an X-ray tube voltage of 120 kV, the coefficient A is 3.704 and B is 0.0367[21,22]. In this study, we calculated four types of WED and four types of SSDE: (1) the mean WED and mean SSDE over the whole scan range; (2) the mean WED and mean SSDE over slices of the lungs, denoted as $WED_{lungs}$ and $SSDE_{lungs}$, respectively; (3) the mean WED and mean SSDE over the slices of the liver, denoted as $WED_{liver}$ and $SSDE_{liver}$, respectively; and (4) the WED and the SSDE at the slice of the middle axial plane outlining the heart, denoted as $WED_{heart\ center}$ and $SSDE_{heart\ center}$, respectively.

### *2.2.2 GPU-based rapid MC simulation*

In this study, we utilized the ARCHER-CT computing engine for rapid MC simulations of CT scans[11]. ARCHER-CT is implemented with a hybrid programming approach using Open Multi-Processing (OpenMP) and Compute Unified Device Architecture (CUDA), supporting CPU parallelism and multi-GPU acceleration. ARCHER-CT is capable of GPU-accelerated simulations of low-energy photon transport below 140 keV in heterogeneous media, modeling photoelectric effects, Compton scattering, and Rayleigh scattering interactions, considering the binding effects of electrons in scatter simulations. As the continuous slowing down approximation (CSDA) range of an electron is generally an order of magnitude smaller than the voxel size on CT, the energy of secondary electrons is presumed to be deposited into local voxels. Through comparison with experimental measurements and the general-purpose MC codes, MCNPX[23], the accuracy of ARCHER-CT has been verified in previous studies[11,24]. ARCHER-CT also supports the specification of multi-detector CT (MDCT) scanner parameters (e.g., photon energy spectrum, bowtie filter shape, fan angle, z-axis collimation, and cone angle settings), making it flexible for modeling various CT scanners. In addition, ARCHER-CT supports Tube Current Modulation (TCM) simulation if the input DICOM file's header contains unique mean mA values for each slice, i.e., mA(z). ARCHER-RT uses mA(z) to determine the specific mA values for longitudinal modulation, thereby adjusting the proportion of particle histories at each projection angle in MC simulations.

In this study, a parameterized and experimentally validated GE LightSpeed Pro 16 MDCT scanner model[17–19] was used to conduct helical scan simulations on the subjects' computational phantoms. Simulated scanning range included the imaged volume without accounting for the Z-over scanning length. The scanning parameters were set to: 120 kVp, beam collimation width 20 mm, body bowtie filter, and pitch 1. A total of 16 X-ray beam sources and bowtie filters were used in a single rotation of helical scan. Since the data collected in this retrospective study were from scans without TCM, the DICOM files did not include mA(z) information, and TCM was therefore not modeled in the MC simulations. The starting angle of the beam in the helical scanning protocols used clinically cannot be obtained from the DICOM file's header; nevertheless, variations in this angle might cause significant discrepancies in the simulated dose for small organs[13,25]. In the MC simulations, we assumed a starting angle of 0°. For each subject, the total number of photon histories used for MC simulation was $10^8$ to maintain the statistical uncertainty (relative standard error, RSE) of most organ doses within 1%, with only the uncertainty of doses to deep and small-volume organs exceeding 1% but less than 2%.

### *2.2.3 Normalized organ dose calculation*

The 3D dose distribution map output by the ARCHER-CT simulation was normalized to each particle and was in voxel-to-voxel correspondence to the input CT image series, in dose units of MeV / (gram·source particle) for each voxel. The organ doses $D_{organ}^{simulated}$, with units of MeV / (gram·source particle), was tallied as follows:

$$D_{organ}^{simulated} = \frac{\sum_i^n D_i^{simulated} \cdot M_i \cdot W_i}{\sum_i^n M_i \cdot W_i} \quad (4)$$

where $i$ is the index of each voxel in the 3D dose matrix $D_i^{simulated}$, and n represents the total number of voxels. $M_i$ is the mass corresponding to voxel $i$, while $W_i$ denotes the value corresponding to voxel $i$ in the organ segmentation mask matrix, which equals 1 if the voxel is within the organ contour and 0 otherwise. To ascertain the CTDI$_{vol}$-normalized organ dose for each subject, it is necessary to multiply the simulated dose by the respective conversion factor (CF), as illustrated in Equation (2):

$$D_{organ}^{normalized}(mGy/CTDI_{vol}^{per\ 100mAs}) = \frac{D_{organ}^{simulated} \cdot CF}{(CTDI_{vol})_{per\ 100mAs}^{validated}} \quad (5)$$

where $D_{organ}^{normalized}(mGy/CTDI_{vol}^{per\ 100mAs})$ represents the organ dose normalized by CTDI$_{vol}$ per 100 mAs, for the subject in units of mGy / (CTDI$_{vol}$ per 100 mAs). $CF$ is the experimentally validated conversion factor for scanning with the GE LightSpeed Pro 16 MDCT device[17,19]. This conversion factor varies with the kVp, beam collimation width and bowtie filter, and is expressed in units of mGy·g·photon / (100 mAs·MeV). CTDI$_{vol}$ is an effective parameter for comparison between different MDCT scanners, and normalization of the organ dose to CTDI$_{vol}$ can eliminate the impact of the CT machine model and scanning parameters. $(CTDI_{vol})_{per\ 100mAs}^{validated}$ is the value derived from previous research in which the GE LightSpeed Pro 16 model was validated against measurements[17,19], and normalized to 100 mAs per rotation. Moreover, previous studies have shown that organ doses normalized by CTDI$_{vol}$ vary by less than 10% across different MDCT scanners[26,27]. However, this study only focuses on CTDI$_{vol}$-normalized organ doses for subjects assumed to have been scanned using a GE LightSpeed Pro 16 MDCT. In other words, the normalized organ doses reported in this manuscript may exhibit slight differences compared to the actual normalized organ doses for subjects scanned with one of the five original scanners.

To facilitate the quantification of the error in estimating patient-specific doses using either the fitted values from a fitted formula or population averages, we calculated the relative error using Equation (6):

$$Relative\ Error = \frac{|OD_{fitted\ or\ mean} - OD_{patient-specific}|}{OD_{patient-specific}} \quad (6)$$

where $OD$ represents the organ dose.

### 2.2.4 Comparison with the ICRP Adult Reference Phantoms

There are discrepancies between individual subjects and population-averaged phantoms, which manifest as differences in organ shape, size, and spatial relationships. When using the population-averaged phantom to represent the corresponding group of individuals, these inconsistencies can lead to errors in organ dose estimates. Furthermore, anatomical differences necessitate meticulous alignment of each subject's CT scan range onto the population-averaged phantom when employing this conventional method. Thus, our study compared the novel patient-specific method with the established ICRP reference phantom method which is a mainstay in radiation protection dosimetry[28]. In the ARCHER-CT simulations, identical scanning parameters were applied to both the ICRP phantoms and the subjects, and the total number of photon histories was $10^{10}$. The scan range chosen for the phantoms closely mirrors the scan range typically employed in our dataset (from C7 to L3).

### 2.2.5 Hardware and software configuration

In this study, both DeepContour and ARCHER-CT were executed on a consumer-grade personal computer (PC) equipped specifically with an Intel Core™ i7-12700 CPU, 64 GB DDR4 memory, and a single NVIDIA GeForce RTX™ 3080 10GB GPU. Statistical analysis was conducted using OriginPro 2021b.

### 2.2.6 Data Availability

Anonymized health metrics, CT scanning parameters, organ masses, organ volumes, and organ doses for the subjects are available from the corresponding author on reasonable request.

## 3 RESULTS

### 3.1 Analysis of collected subject health metrics

Tables 1 presents the statistics on age, gender, and body metrics for the 10,281 subjects. There were 6419 male subjects, with an average (± standard deviation) age (years), height (cm), weight (kg), BMI (kg/m$^2$), average water equivalent diameter (WED) (cm) and scan length (cm) of 49.5 (±12.7), 170.6 (±6.1), 73.8 (±9.9), 25.3 (±2.9), 26.2 (±1.8), and 36.1 (±2.2), respectively. Among the 3862 female subjects, the average age, height (cm), weight (kg), BMI (kg/m$^2$), mean WED (cm) and scan length (cm) were 49.7 (±12.9), 159.1 (±5.7), 58.8 (±8.2), 23.2 (±3.1), 23.9 (±2.0), and 34.3 (±1.9), respectively.

**Table 1.** Age, gender, and body metric statistics for the 10,281 subjects: 6419 males and 3862 females.

| Gender | Metric | Age | Height (cm) | Weight (kg) | BMI (kg/m$^2$) | Average WED (cm) | Scan length (cm) |
|---|---|---|---|---|---|---|---|
| Male | Average (±standard deviation) | 49.5 (±12.7) | 170.6 (±6.1) | 73.8 (±9.9) | 25.3 (±2.9) | 26.2 (±1.8) | 36.1 (±2.2) |
|  | Range | 19~93 | 146.0~195.0 | 44.3~136.2 | 15.8~41.9 | 19.8~34.5 | 29.5~46.3 |
| Female | Average (±standard deviation) | 49.7 (±12.9) | 159.1 (±5.7) | 58.8 (±8.2) | 23.2 (±3.1) | 23.9 (±2.0) | 34.3 (±1.9) |
|  | Range | 16~93 | 139.0~187.0 | 36.5~111.2 | 15.8~44.1 | 18.4~34.4 | 25.3~42.1 |

Table 2 and Figure 2 summarizes the data of the 10,281 subjects according to the BMI categories established globally by the World Health Organization (WHO)[29]. Among all the subjects, 56.6% were classified in the normal weight category, while 37.5% were classified as overweight. The other BMI categories accounted for 5.9% of the subjects. Notably, the BMI distribution differed between the sexes. Among males, the proportions of individuals with a normal weight and overweight were similar; in contrast, in the female cohort, 73% of the subjects were in the normal weight subgroup, whereas 21% were in the overweight subgroup.

**Table 2.** BMI data of the 10,281 subjects according to WHO BMI category[29].

| Category | BMI Classification standard (kg/m$^2$) | Male subjects | Average (± standard deviation) BMI of the male subjects (kg/m$^2$) | Female subjects | Average (± standard deviation) BMI of the female subjects (kg/m$^2$) |
|---|---|---|---|---|---|
| Underweight | (0, 18.5) | 40 | 17.9 (±0.7) | 120 | 17.8 (±0.6) |
| Normal weight | [18.5, 25) | 2998 | 23.1 (±1.4) | 2819 | 22.1 (± 1.7) |
| Overweight | [25, 30) | 3037 | 26.9 (± 1.3) | 815 | 26.8 (± 1.3) |
| Obese Ⅰ | [30, 35) | 314 | 31.4 (± 1.1) | 97 | 31.5 (± 1.3) |
| Obese Ⅱ | [35, 40) | 27 | 36.7 (± 1.4) | 10 | 36.8 (± 1.1) |
| Morbidly obese | [40, ∞) | 3 | 41.2 (± 0.8) | 1 | 44.1 |

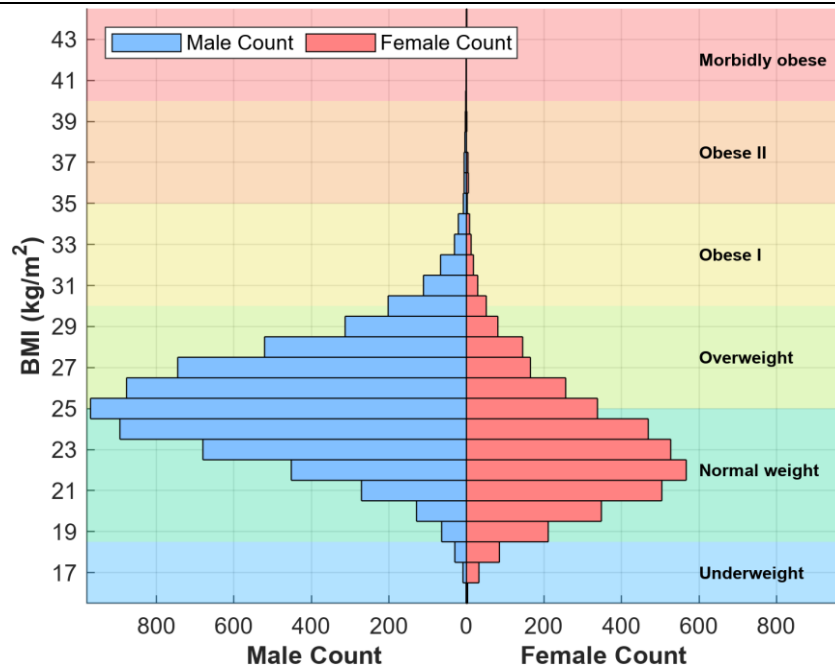

**Figure 2.** Histogram of the number of subjects as a function of BMI, where red bars represent female subjects and blue bars represent male subjects. BMI categories are represented by background colors.

The data covers scans from five different CT scanner models: GE Discovery CT750 HD, GE LightSpeed VCT, NMS NeuViz 128, GE Optima CT660, and GE Optima CT680 Series. Notably, for all the models, the kilovoltage peak (kVp) was uniformly set at 120, with a collimation width of 40 mm. All the models utilized the body-type bowtie filter.

As shown in Table 3, we performed linear fitting for the four types of WED and BMI. The results indicate that BMI is most effective for estimating the mean WED over the scan range.

**Table 3.** Linear fitting formulas for the four types of WED and BMI for males and females

| Target | | Mean WED | $WED_{lungs}$ | $WED_{liver}$ | $WED_{heart\ center}$ |
|---|---|---|---|---|---|
| Male | Formula | 0.591*BMI+11.26 | 0.589*BMI+10.85 | 0.657*BMI+10.98 | 0.607*BMI+9.03 |
| | $R^2$ | 0.85 | 0.83 | 0.81 | 0.78 |
| Female | Formula | 0.602*BMI+9.88 | 0.602*BMI+9.58 | 0.620*BMI+10.07 | 0.609*BMI+8.35 |
| | $R^2$ | 0.84 | 0.82 | 0.81 | 0.75 |

## 3.2 Patient-specific organ dose database

For each CT scan subject, the average simulation time for ARCHER-CT was 10.2 (±2.2) seconds, and the mean time for DeepContour's preprocessing and automatic segmentation was approximately 124.7 (±14.3) seconds.

Figure 3 presents a visualization of patient-specific phantom examples constructed from a subject's CT data using

DeepContour and the corresponding dose distribution simulated by ARCHER-CT.

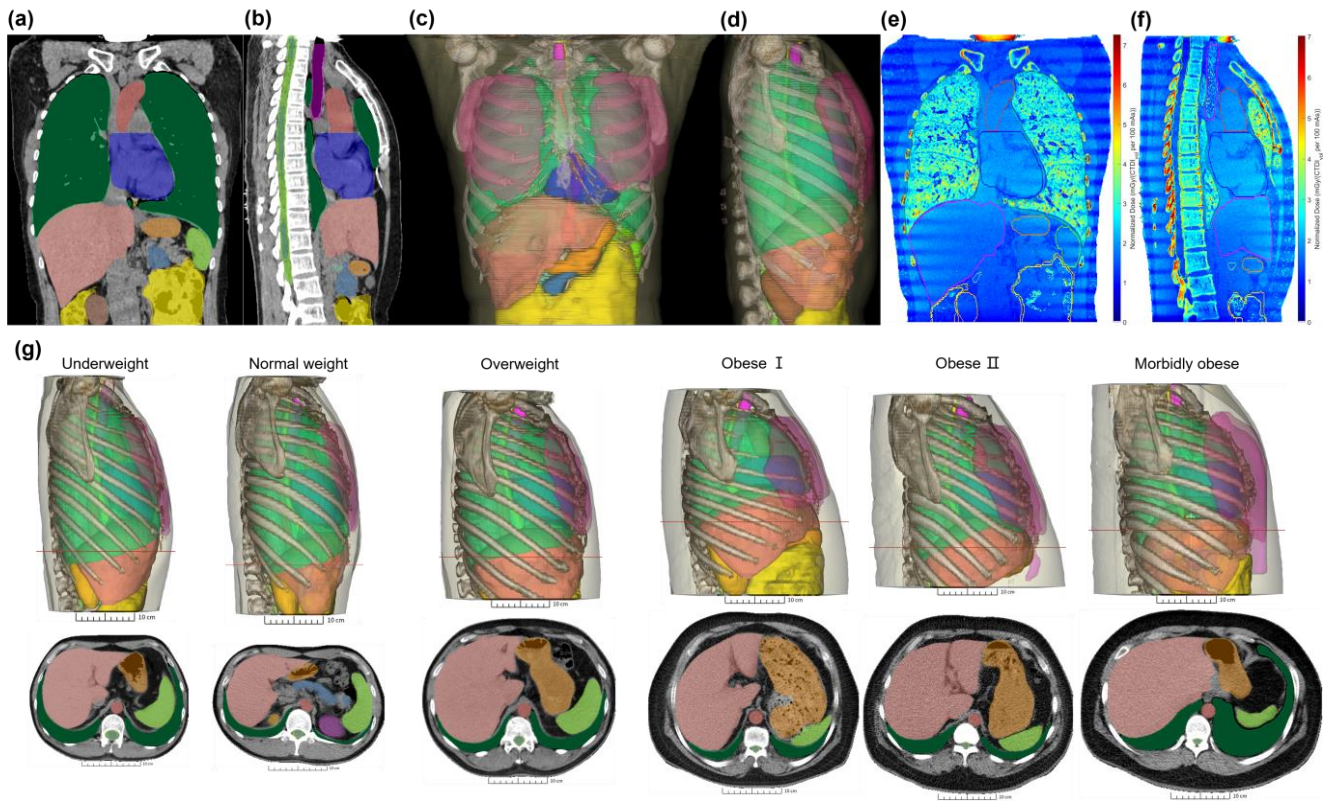

**Figure 3.** Visualization of patient-specific phantom examples constructed from CT data and multi-organ segmentation, as well as the corresponding dose distribution. (a)(b) Slices of the CT image and automatic segmentation masks. (c)(d) views of the 3D geometric phantom. (e)(f) Slices of the $CTDI_{vol}$-normalized dose distribution computed by MC simulation. (a)-(f) Example phantom for a female subject. (g) Example patient-specific male phantoms for six different BMI categories as defined by the World Health Organization [29], from underweight to morbidly obese, illustrating differences in organ shape, size and spatial relationships among the categories.

Figure 4 illustrates the distribution of the patient-specific $CTDI_{vol}$-normalized organ doses, as well as the results from ICRP Adult Reference Phantoms. In males, the patient-specific doses to the esophagus, heart, lungs, liver, pancreas, spleen, stomach, breasts, and spinal cord are found to range between 60% to 134%, 67% to 141%, 80% to 153%, 55% to 125%, 33% to 128%, 49% to 121%, 51% to 126%, 81% to 143%, 68% to 154%, respectively, normalized by the doses from the ICRP reference phantom. In females, the patient-specific doses to the nine organs are found to range between 52% to 126%, 53% to 120%, 64% to 140%, 47% to 109%, 37% to 119%, 56% to 107%, 49% to 110%, 72% to 121%, 71% to 164%, respectively, normalized by the doses from the ICRP reference phantom method.

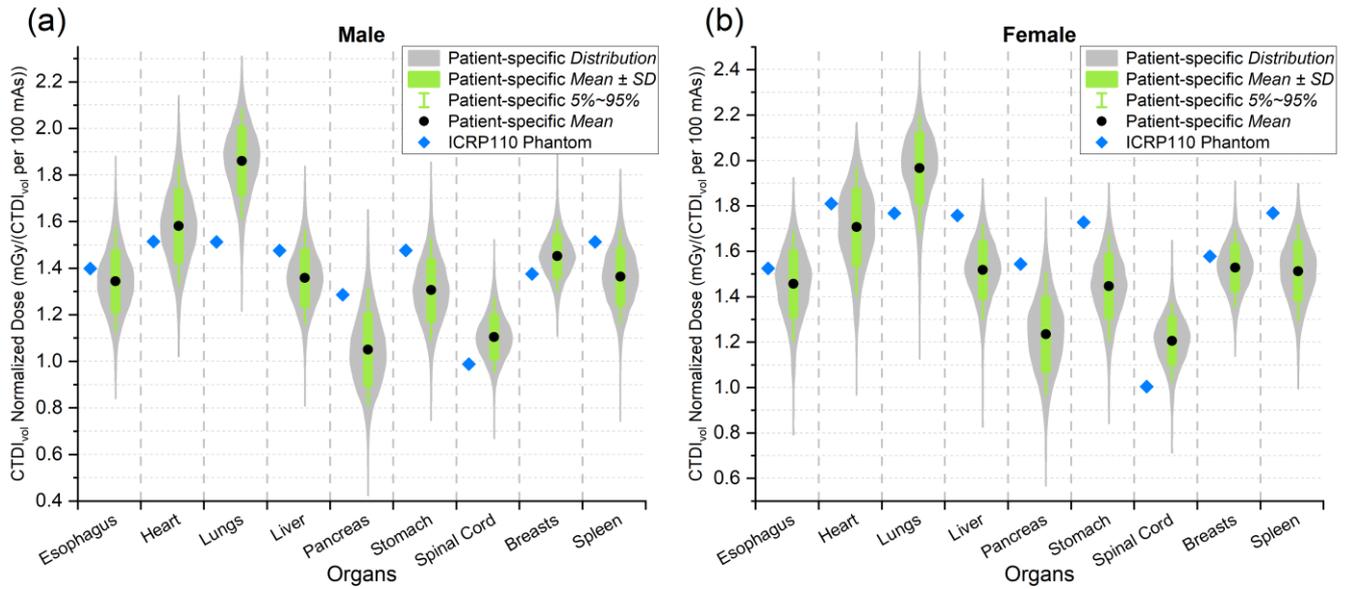

**Figure 4.** Comparison of the proposed patient-specific method with the results from ICRP Adult Reference Phantoms[28]. The vertical axis displays the CTDI$_{vol}$ normalized organ dose in units of mGy / (CTDI$_{vol}$ per 100 mAs). The horizontal axis shows the nine organs. (a) Results for males. (b) Results for females.

Furthermore, Table 4 provides the average values and ranges of organ doses for six different mean WED intervals obtained from this dataset. Considering both sexes and organs except for the pancreas, the relative error of the averages of organ doses with respect to the patient-specific values reached up to 19%, 30%, 37%, 42%, 30%, and 49%, respectively, for the six mean WED groups. When the pancreas doses of 14 subjects were estimated using the average of the population with similar WED, the relative error exceeded 50%, with a maximum of 77%. This suggests that WED-based average phantoms are not as accurate when compared with patient-specific phantoms, and the range of such errors can be found in Table 4. It is obvious that, organ density, volume, and anatomical structure are all indeed different when a big-enough population data such as that the one used in this study is examined.

**Table 4.** Mean (minimum~maximum) CTDI$_{vol}$-normalized doses (mGy / (CTDI$_{vol}$ per 100 mAs)) of the nine organs for six different mean WED intervals

| Mean WED interval (cm) | [18.4, 21] | | (21,23.5] | | (23.5,26] | | (26,28.5] | | (28.5,31] | | (31,34.5] | |
|---|---|---|---|---|---|---|---|---|---|---|---|---|
| Gender | Male | Female | Male | Female | Male | Female | Male | Female | Male | Female | Male | Female |
| Spinal cord dose | 1.42 (1.31~1.52) | 1.38 (1.22~1.65) | 1.28 (1.08~1.50) | 1.27 (1.10~1.51) | 1.16 (0.99~1.40) | 1.17 (0.95~1.35) | 1.06 (0.91~1.24) | 1.07 (0.92~1.21) | 0.97 (0.84~1.15) | 0.95 (0.88~1.05) | 0.86 (0.67~1.05) | 0.83 (0.71~0.97) |
| Breasts dose | 1.71 (1.55~1.90) | 1.70 (1.49~1.91) | 1.61 (1.42~1.97) | 1.59 (1.42~1.79) | 1.51 (1.31~1.79) | 1.49 (1.29~1.68) | 1.41 (1.22~1.60) | 1.39 (1.25~1.50) | 1.32 (1.15~1.45) | 1.29 (1.19~1.40) | 1.23 (1.11~1.33) | 1.20 (1.14~1.26) |
| Esophagus dose | 1.73 (1.60~1.88) | 1.71 (1.50~1.92) | 1.59 (1.40~1.79) | 1.55 (1.31~1.82) | 1.43 (1.25~1.69) | 1.40 (1.14~1.60) | 1.28 (1.09~1.50) | 1.24 (1.04~1.39) | 1.14 (0.95~1.30) | 1.09 (0.98~1.20) | 0.97 (0.84~1.09) | 0.94 (0.79~1.02) |
| Heart dose | 2.05 (1.95~2.14) | 1.99 (1.83~2.17) | 1.88 (1.64~2.06) | 1.82 (1.52~2.04) | 1.68 (1.45~1.92) | 1.64 (1.37~1.86) | 1.51 (1.25~1.75) | 1.46 (1.25~1.65) | 1.34 (1.16~1.50) | 1.30 (1.16~1.42) | 1.16 (1.02~1.30) | 1.13 (0.97~1.24) |
| Lungs dose | 2.16 (2.08~2.25) | 2.17 (1.95~2.48) | 2.07 (1.85~2.31) | 2.05 (1.67~2.42) | 1.95 (1.64~2.25) | 1.93 (1.53~2.29) | 1.81 (1.38~2.13) | 1.77 (1.38~2.05) | 1.65 (1.30~1.92) | 1.59 (1.27~1.79) | 1.45 (1.21~1.66) | 1.38 (1.13~1.63) |
| Liver dose | 1.70 (1.57~1.82) | 1.71 (1.44~1.92) | 1.56 (1.34~1.84) | 1.59 (1.30~1.83) | 1.43 (1.15~1.73) | 1.48 (1.22~1.72) | 1.31 (1.07~1.54) | 1.36 (1.13~1.55) | 1.19 (1.01~1.42) | 1.21 (0.93~1.37) | 1.03 (0.81~1.19) | 1.04 (0.83~1.17) |
| Pancreas dose | 1.42 (1.10~1.65) | 1.44 (1.13~1.84) | 1.28 (0.89~1.65) | 1.30 (0.83~1.73) | 1.12 (0.69~1.55) | 1.20 (0.83~1.52) | 0.99 (0.61~1.35) | 1.08 (0.72~1.37) | 0.88 (0.59~1.19) | 0.95 (0.61~1.21) | 0.75 (0.42~0.97) | 0.77 (0.57~0.97) |
| Stomach dose | 1.65 (1.43~1.86) | 1.63 (1.39~1.90) | 1.51 (1.27~1.76) | 1.51 (1.16~1.81) | 1.38 (1.00~1.64) | 1.42 (1.09~1.68) | 1.26 (0.92~1.52) | 1.28 (0.91~1.53) | 1.13 (0.92~1.31) | 1.15 (0.92~1.32) | 0.98 (0.75~1.15) | 1.00 (0.84~1.12) |
| Spleen dose | 1.68 (1.48~1.80) | 1.67 (1.45~1.90) | 1.54 (1.24~1.82) | 1.57 (1.21~1.88) | 1.42 (1.10~1.75) | 1.48 (1.11~1.79) | 1.32 (1.03~1.61) | 1.38 (1.06~1.63) | 1.24 (0.97~1.49) | 1.28 (1.08~1.52) | 1.11 (0.74~1.28) | 1.12 (0.99~1.26) |
| Sample size | 10 | 241 | 422 | 1535 | 2447 | 1554 | 2885 | 457 | 606 | 67 | 49 | 8 |

## 3.3 Statistical analysis of organ doses and health metrics

Notably, in our dataset of 10,500 original cases, 81 subjects underwent repeated CT examinations. Studies have shown that subjects who receive multiple CT scans have a considerable probability of surviving more than ten years after the examinations, during which they may face radiation-induced risks[30]. Furthermore, the advantage of having such a big-data information is that it enables the analysis of organ doses for a large population of subjects, thus addressing the critical need for potentially more accurate radiation risk assessment[31]. To this end, we herein provide extensive big-data statistical analysis using organ dose data summarized in the last section.

Figure 5 presents the statistical distribution of doses normalized by $CTDI_{vol}$ for nine organs relative to the mean WED, as well as exponential fit lines. As can be seen, organ dose decreases as WED increases, consistent with previous studies. The exponential fit curves lie centrally within the scattered points, providing an average estimate for each WED value based on the fitting formula. However, only the doses of heart and esophagus exhibit an $R^2$ (the coefficient of determination) value around 0.9 for the fitted curves, while for other organ doses, the $R^2$ values are below 0.8. Moreover, for any given WED value in the scatter plot, there is noticeable variation in organ doses among subjects, with a wide range of values. For a given WED (accurate to one decimal place), there can be as much as 32%, 29%, 47%, 41%, 113%, 56%, 67%, 32%, and 40% variation, respectively in the dose of the esophagus, heart, lungs, liver, pancreas, spleen, stomach, breasts, and spinal cord. Considering both sexes, the relative error of the organ doses from fitted formulas with respect to the patient-specific values reached up to 18%, 16%, 32%, 37%, 59%, 40%, 40%, 17%, and 20%, respectively for the nine organs.

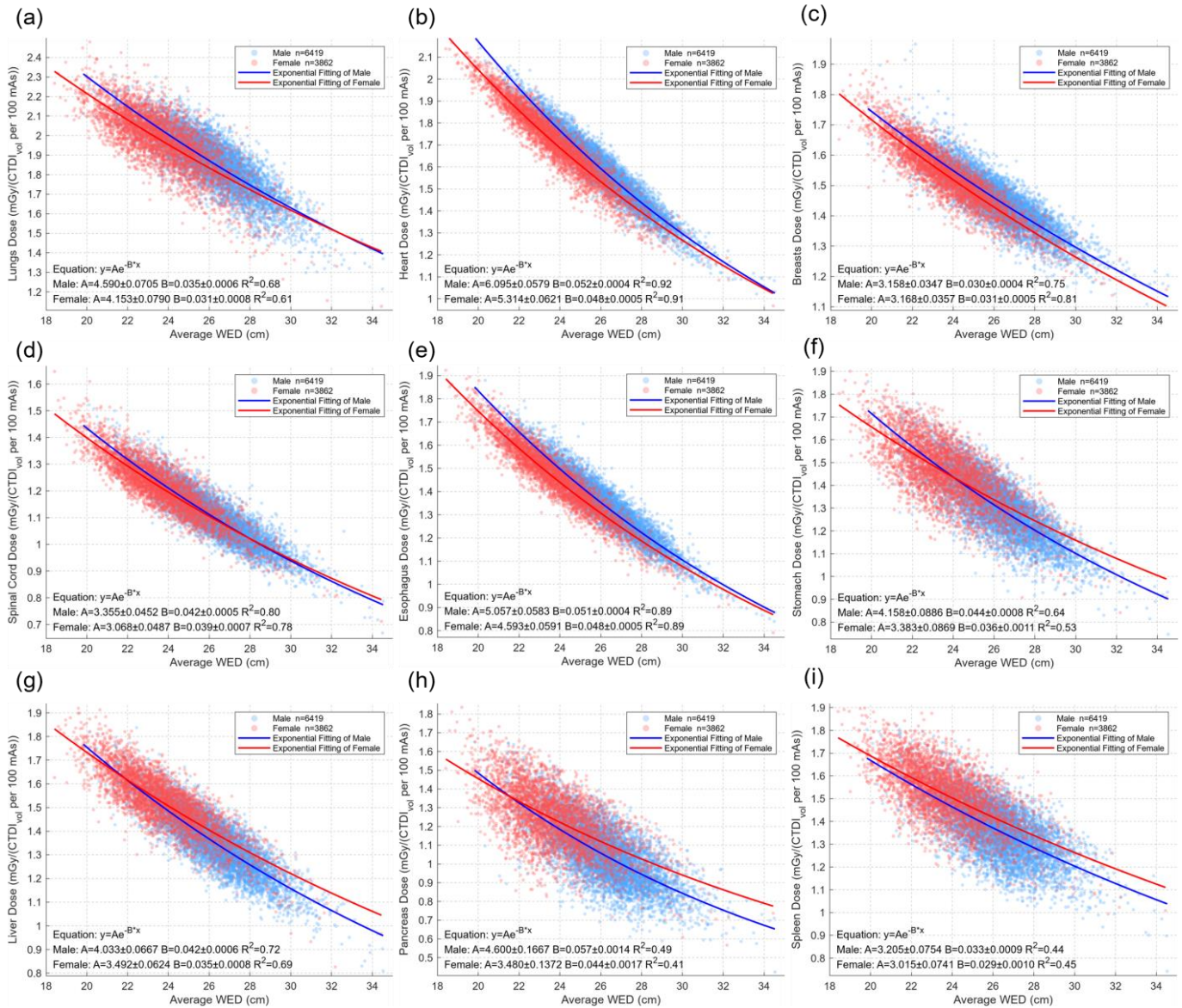

**Figure 5.** Graphs of the $CTDI_{vol}$-normalized organ doses in mGy / (CTDIvol per 100 mAs) calculated from CT scans for 9 organs relative to the mean WED, including exponential fit curves. (a)-(i) Results for the lungs, heart, breasts, spinal cord, esophagus, stomach, liver, pancreas, and spleen, respectively.

Many believe that, in radiation protection dosimetry, the tolerance level for error is 50%, then it is worth noting that, our big-data analysis suggests something profound — i.e., many of previous CT organ dosimetry and understanding of associated risk using "average" phantom methods may be too uncertain to be meaningful.

Furthermore, we analyzed the sensitivity of number of CT dosimetry parameters using Pearson correlation method. As shown in Figure 6, we calculated Pearson correlation coefficients to investigate the relationship between $CTDI_{vol}$-normalized organ doses and various health metrics, including gender (coded as 1 for male and 0 for female), age at exposure, height, weight, BMI, scan length, and the four types of SSDE. These factors provide information on how to select CT scan settings in order to minimize the impact on organ doses.

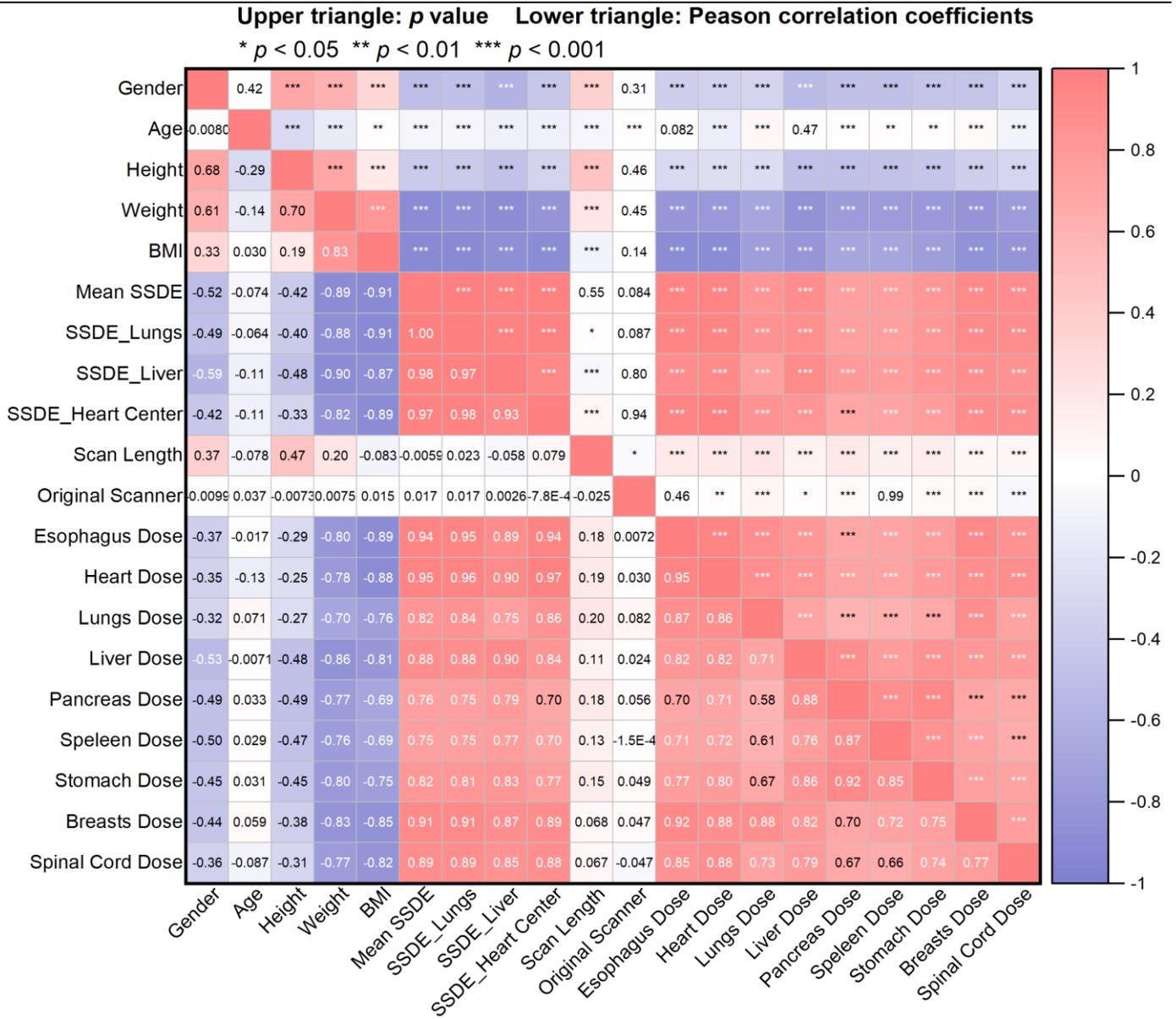

**Figure 6.** Pearson correlation coefficients between the $CTDI_{vol}$-normalized organ doses for CT scans and other health metrics for the subjects.

After analyzing the Pearson correlation coefficients among the quantities, we identified the following five interesting findings:

1) There is a strong correlation between gender and both height and weight (absolute values > 0.6). In contrast, the correlation between gender and BMI is weaker (absolute value = 0.33).

2) Weight and BMI have very strong correlations with the four types of SSDE (absolute values > 0.8).

3) The four types of SSDE are highly correlated with each other (values > 0.9). Notably, the correlation between mean SSDE and $SSDE_{lungs}$ is nearly perfect (value = 1), suggesting they are nearly equivalent within this dataset.

4) Lungs and heart doses correlate most strongly with $SSDE_{heart\ center}$ (values of 0.86 and 0.97, respectively), indicating $SSDE_{heart\ center}$ as a reliable surrogate for normalized doses of lungs and heart. Other thoracic organ doses can be best predicted by $SSDE_{lungs}$ or mean SSDE. Abdominal organ doses are most strongly correlated with SSDE$_{liver}$ (values between 0.77 and 0.90), suggesting SSDE$_{liver}$ as the best surrogate for abdominal organ doses.

5) When WED information is unavailable and thus SSDE cannot be calculated, BMI correlates most strongly with thoracic organ doses (absolute values between 0.76 and 0.89), while weight correlates most strongly with abdominal organ doses (absolute values between 0.77 and 0.86).

## 4. Discussion

The patient-specific method proposed in this study shows high accuracy and cost-effectiveness, offering the potential to enable more informed decision-making regarding radiation exposure, thereby enhancing the safety and efficacy of clinical imaging diagnostics. On average, it took 135 seconds per subject to compute all organ doses. In other words, using a single NVIDIA RTX 3080 GPU card, this clinically deployable method can automatically evaluate patient-specific organ doses for over 600 subjects per day, without requiring manual intervention. Furthermore, integrating this method into clinical systems to conduct big data analysis could facilitate data mining and trend prediction of quality control and radiation risk.

This study has some limitations. First, due to the data collected from this retrospective study using fixed tube current, we did not consider Automatic Exposure Control (AEC) technologies, such as Automated Tube Current Modulation (ATCM). Second, although a validated model of a 16-detector CT scanner was used in this study, MDCT with more detector slices is becoming increasingly common in clinical settings. Scanner models that better represent current machines should be developed and validated. Additionally, this study only considered the CT scan range of the lower neck, thorax, and upper abdomen, meaning that some of our conclusions may not be applicable to CT scans of other regions. Lastly, our simulations considered only the imaged region and did not account for the effects of Z-over scanning length or scattered radiation from outside the scan range on organ doses, thus limiting the accuracy of dose estimates for organs partially outside the scanned volume.

## 5. Conclusions

This study has demonstrated a state-of-the-art accurate and rapid MC patient-specific organ-level dose estimation method involving automatic segmentation of, for the first time, a cohort of 10,281 subjects undergoing CT examinations as part of a workflow in a busy Chinese hospital. We have derived a large organ dose dataset that reveals surprisingly wide range of dose variation among the individuals we studied. We compared our data with those reported in the literature to show that population-averaged phantoms such as those recommended by the ICRP Reference Man concept, can lead to considerable dosimetric uncertainties. The results presented in this study made us agree with suggestions in the literature that previous CT organ dose data and understanding of associated risk based on old and obsolete dosimetry methods can and should be updated[32,33].

# Acknowledgement

The study is supported by the following funding sources: Anhui Province Key Research and Development Plan (2023s07020020), National Natural Science Foundation of China (12275372), USTC Fund for the Development of Medical Physics and Biomedical Engineering Interdisciplinary Subjects (YD2140000601), and Research Funds of Joint Research Center for Regional Diseases of IHM (2023bydjk002).
# Conflict of interests

None of the authors report any conflicts of interest relevant to this article.

## References

1. United Nations: Scientific Committee on the Effects of Atomic Radiation. *UNSCEAR 2020/2021 Report Volume I, Scientific Annex A: 'Medical Exposure to Ionizing Radiation.'* United Nations; 2022. 354p.

2. National Council on Radiation Protection and Measurements. *Medical Radiation Exposure of Patients in the United States: Recommendations of the National Council on Radiation Protection and Measurements*. NCRP; 2019. 285 p.

3. Yao J, Gao L, Qian A, Wang B. Survey on frequency of medical X-ray diagnosis in Shanghai. *Chin J Radiol Med Prot*. 2019;39(5):370-375. doi:10.3760/CMA.J.ISSN.0254-5098.2019.05.009

4. Bosch de Basea Gomez M, Thierry-Chef I, Harbron R, et al. Risk of hematological malignancies from CT radiation exposure in children, adolescents and young adults. *Nat Med*. 2023;29:3111-3119. doi:10.1038/s41591-023-02620-0

5. National Research Council. *Health Risks from Exposure to Low Levels of Ionizing Radiation: BEIR VII Phase 2*. The National Academies Press; 2006. 422 p.

6. Rogers DWO. Fifty years of Monte Carlo simulations for medical physics. *Phys Med Biol*. 2006;51(13):R287-R301. doi:10.1088/0031-9155/51/13/R17

7. Källman HE, Traneus E, Ahnesjö A. Toward automated and personalized organ dose determination in CT examinations — A comparison of two tissue characterization models for Monte Carlo organ dose calculation with a Therapy Planning System. *Med Phys*. 2019;46(2):1012-1023. doi:10.1002/mp.13357

8. Shi F, Hu W, Wu J, et al. Deep learning empowered volume delineation of whole-body organs-at-risk for accelerated radiotherapy. *Nat Commun*. 2022;13(1):6566. doi:10.1038/s41467-022-34257-x

9. Wasserthal J, Breit HC, Meyer MT, et al. TotalSegmentator: Robust Segmentation of 104 Anatomic Structures in CT Images. *Radiol Artif Intell*. 2023;5(5):e230024. doi:10.1148/ryai.230024